\documentclass[aps,pre,twocolumn,superscriptaddress,preprintnumbers]{revtex4-2}

\usepackage{amsmath,amssymb}
\usepackage{graphicx}
\usepackage{dcolumn}
\usepackage{bm}
\usepackage{color}
\usepackage{hyperref}
\usepackage{physics}

\begin{document}

%--------------------------------------------------------------------------
% TITLE
%--------------------------------------------------------------------------
\title{Complex spacing ratio statistics in the partially open asymmetric quantum baker map}

\author{Leonardo Ermann}
\affiliation{Dto.~FTIFSC, GIyA, Comisi\'on Nacional de Energ\'ia At\'omica (CNEA),
Av.~del Libertador 8250, Buenos Aires 1429, Argentina}
\affiliation{Dto.~FTIFSC, GIyA, Consejo Nacional de Investigaciones Cient\'ificas y T\'ecnicas (CONICET), Argentina}

\author{Pablo Sesin}
\affiliation{Dto.~FTIFSC, GIyA, Comisi\'on Nacional de Energ\'ia At\'omica (CNEA),
Av.~del Libertador 8250, Buenos Aires 1429, Argentina}

\author{Alejandro M. F. Rivas}
\affiliation{Dto.~FTIFSC, GIyA, Comisi\'on Nacional de Energ\'ia At\'omica (CNEA),
Av.~del Libertador 8250, Buenos Aires 1429, Argentina}
\affiliation{Dto.~FTIFSC, GIyA, Consejo Nacional de Investigaciones Cient\'ificas y T\'ecnicas (CONICET), Argentina}

\author{Pablo Bergamasco}
\affiliation{Dto.~FTIFSC, GIyA, Comisi\'on Nacional de Energ\'ia At\'omica (CNEA),
Av.~del Libertador 8250, Buenos Aires 1429, Argentina}

\author{Gabriel G. Carlo}
\affiliation{Dto.~FTIFSC, GIyA, Comisi\'on Nacional de Energ\'ia At\'omica (CNEA),
Av.~del Libertador 8250, Buenos Aires 1429, Argentina}
\affiliation{Dto.~FTIFSC, GIyA, Consejo Nacional de Investigaciones Cient\'ificas y T\'ecnicas (CONICET), Argentina}

%--------------------------------------------------------------------------
% ABSTRACT
%--------------------------------------------------------------------------
\begin{abstract}
We study the complex eigenvalue statistics of the asymmetric quantum baker map
with partial projective openings. The classical asymmetric baker map, with its
discontinuity at $q=2/3$, is fully chaotic, has no reflection symmetry, and
provides a clean setting with tunable escape rate and fractal repeller dimension.
We consider three distinct opening geometries in position space:
localized (contiguous channels), random, and uniform (equispaced channels),
all controlled by a tunable amplitude reflectivity parameter $\rho$ that interpolates
between the fully open ($\rho=0$) and the closed ($\rho=1$) limits.
We use the partially truncated circular unitary ensemble (PTCUE) as the 
random matrix theory benchmark. 
The main focus is on the joint distribution of the complex spacing ratio $z$,
defined as the ratio of the distances from an eigenvalue to its nearest and
next-nearest neighbors in the complex plane. We find a smooth crossover from
a quasi-1D spectral regime, where eigenvalues cluster near the unit circle and
the phase distribution of $z$ is peaked, to a two-dimensional Ginibre-like
regime, where the distribution becomes nearly uniform and level repulsion is
fully developed. Both the number of open channels $M$ and the reflectivity $\rho$
modulate this crossover, and $\rho$ provides an additional continuous 
control even at fixed opening size. 
All three opening models converge to PTCUE statistics at large $M$,
while differences are most pronounced for the localized model at small $M$.
No evidence of an abrupt transition is found. This crossover which suggests a 
universal behavior, has deep consequences for open quantum and 
wave-chaotic experiments.
\end{abstract}


\maketitle

%--------------------------------------------------------------------------
% I. INTRODUCTION
%--------------------------------------------------------------------------
\section{Introduction}
\label{sec:intro}

When a quantum system with chaotic classical dynamics is coupled to the outside
world, its spectrum migrates into the complex plane. The eigenvalues of the
non-unitary propagator become resonances, characterized by a finite lifetime
set by how quickly probability leaks through the opening.
Understanding the statistical properties of these resonances is a central problem
in quantum chaos \cite{haake19,altmann13,novaes13}.

For fully chaotic open systems, the Grobe-Haake-Sommers (GHS) conjecture states
that the resonance spectrum should follow the statistics of complex Ginibre random
matrices \cite{grobe88,ginibre65,hamazaki20}. Extensive numerical evidence has
supported this picture in a variety of systems, including quantum maps with
classical openings (holes that remove all probability on arrival)
\cite{novaes13,altmann13}. In this regime the relevant random matrix model is the
truncated circular unitary ensemble (TCUE), in which some columns of a CUE matrix
are set to zero to mimic projective leakage \cite{novaes13}. The fractal Weyl law
relates the number of long-lived resonances to the fractal dimension of the
classical repeller \cite{novaes13,korber13,schoenwetter15,nonnenmacher12,schmidt23}.

The baker map and its relatives have been particularly useful as model systems to
study open quantum chaos, owing to their exact symbolic dynamics and simple
quantization \cite{balazvoros89,saraceno90,ermann06baker}. Resonance eigenstates
of open baker maps have been shown to localize on the classical repeller
\cite{ermann09prl}, and the transient features of the resonance spectrum have been
investigated for different opening geometries \cite{ermann12prl}. In the latter
context it was found that the fine details of the resonance distribution are
sensitive to the specific shape of the opening even when the asymptotic properties
(repeller dimension, escape rate) are the same. The partially open tribaker map,
studied semiclassically with short periodic orbit theory \cite{carlo16pre,prado18},
showed that the transition from qualitatively open to qualitatively closed behavior
can be traced to orbits that are not part of the classical repeller.

A distinct and experimentally relevant class of systems are those with
\textit{partial} leakage: optical microcavities, microwave billiards, and
semiconductor quantum dots all have boundaries with finite reflectivity, so that
an incoming trajectory is only partly absorbed and partly reflected back
\cite{cao15,gmachl98,altmann13}. In these systems, the resonance statistics are
crucially shaped by the partial opening at the interface, as seen in the
formation of whispering-gallery modes and chaotic directional emission in
asymmetric microcavities \cite{wiersig08,shinohara10}.
In the classical limit this is described by a
leaking chaotic system with partial absorption \cite{altmann13,schoenwetter15}.
Quantum mechanically, multiplying the columns of the unitary propagator by a
factor $\rho\in[0,1]$ (amplitude reflectivity) instead of setting them to zero
\cite{carlo16pre} generates a family of partially open maps that interpolate
between the fully open ($\rho=0$) and the closed ($\rho=1$) regimes.

While the density of states and fractal Weyl law for partially open maps have
been studied \cite{carlo16pre,prado18,schoenwetter15}, much less is known about
the spectral fluctuations and level statistics in this partial-opening regime.
Very recently, Signor et al. \cite{santos26} studied the spectral correlations
of the leaky quantum standard map with a localized opening in detail and found
that the system does not follow unconstrained Ginibre statistics; instead,
the short-range correlations are governed by the non-Hermitian symmetry class
AI$^\dagger$. They also showed that as the opening size grows, the density
of states of the truncated ensemble approaches the Ginibre circular law.
Their work highlighted that the type of opening (localized vs.\ extended) matters
for spectral correlations, and motivated a more detailed study of the full
crossover between different statistical regimes.

In this paper we study the spectral fluctuations of the asymmetric quantum baker
map with three types of partial projective openings across a wide range of
opening sizes $M$ and reflectivities $\rho$. The asymmetric baker map, with its
classical discontinuity at $q=2/3$, is fully chaotic and has no reflection
symmetry in phase space; this avoids the degeneracies that affect the symmetric
baker map and makes it an ideal testbed for statistical studies
\cite{ermann06baker,ermann08asymbaker}. We introduce a reflectivity
parameter $\rho$ and three opening geometries: localized (contiguous block), random,
and uniform (equispaced).

As the main observable we use the complex spacing ratio \cite{sa20}
\begin{equation}
z_i = \frac{\lambda_{\mathrm{NN},i} - \lambda_i}{\lambda_{\mathrm{NNN},i} - \lambda_i},
\label{eq:z}
\end{equation}
where $\lambda_{\mathrm{NN},i}$ and $\lambda_{\mathrm{NNN},i}$ are the nearest
and next-nearest neighbors of $\lambda_i$ in the complex plane.
This quantity does not require unfolding and gives access to the joint distribution
$P(|z|,\theta_z)$ of the modulus $|z|$ and the phase $\theta_z = \arg(z)$.
While originally introduced for dissipative spin chains \cite{sa20}, this
metric has recently seen a surge in interest as a tool to identify non-Hermitian
universality classes and Ginibre-like chaos in a variety of many-body dissipative
and non-Hermitian systems \cite{garcia23,sa23}.
This is more informative than marginals alone because it captures the 2D structure
of the spectral correlations in the complex plane.

Our main findings are: (i) for small $M$ channels and/or large $\rho$, the spectrum is
quasi-unitary and the $z$ statistics are effectively one-dimensional, with peaks
in $P(\theta_z)$ at $\theta_z=0$ and $\theta_z=\pm\pi$; (ii) as $M$ increases or
$\rho$ decreases toward zero, there is a smooth crossover to 2D Ginibre-like statistics
with a nearly uniform $P(\theta_z)$ and a growing power-law exponent $\beta$ in
$P(|z|)\sim|z|^\beta$; (iii) the reflectivity $\rho$ provides a universal
route to the crossover, accessible even for a single open channel; (iv) all
three opening models converge to partially truncated circular unitary ensemble 
(PTCUE) at large $M$, while differences are most
pronounced at small $M$, and they are well captured by the Bhattacharyya overlap
of the 2D $z$ distributions; (v) the crossover is smooth with no evidence of an
abrupt transition for any combination of $M$ and $\rho$.

The paper is organized as follows. In Sec.~\ref{sec:model} we describe the
classical and quantum asymmetric baker map with partial openings.
Section~\ref{sec:rmt} introduces the PTCUE benchmark and the complex spacing
ratio $z$, and presents the eigenvalue and $z$ distributions for representative
cases (Fig.~\ref{fig:2}). Section~\ref{sec:evol} studies the
evolution of the spectral statistics with $M$ and $\rho$ (Figs.~\ref{fig:3},
\ref{fig:4}, and \ref{fig:5}). Section~\ref{sec:compare} quantifies the similarity
between models using the Bhattacharyya coefficient (Fig.~\ref{fig:6}).
Conclusions are presented in Sec.~\ref{sec:conclusions}.

%--------------------------------------------------------------------------
% II. THE MODEL
%--------------------------------------------------------------------------
\section{The model}
\label{sec:model}

\subsection{Classical asymmetric baker map}
\label{sec:classical}

The classical system is the baker map on the unit torus
$\mathbb{T}^2=[0,1)\times[0,1)$, with a discontinuity at $q=\alpha=2/3$:
\begin{equation}
\mathcal{B}(q,p) = \begin{cases}
\left(\dfrac{q}{\alpha},\; \alpha\, p\right) & 0 \le q < \alpha, \\[4pt]
\left(\dfrac{q-\alpha}{1-\alpha},\; (1-\alpha)\,p + \alpha\right) & \alpha \le q < 1.
\end{cases}
\label{eq:baker_classical}
\end{equation}
This map preserves the Lebesgue measure on $\mathbb{T}^2$ and is fully chaotic (see, e.g., \cite{katok1995}).
Unlike the symmetric baker ($\alpha=1/2$), the map at $\alpha=2/3$ has two distinct
local Lyapunov exponents, one per branch:
\begin{align}
\Lambda_1 &= \ln\frac{1}{\alpha} = \ln\tfrac{3}{2} \approx 0.405, \nonumber\\
\Lambda_2 &= \ln\frac{1}{1-\alpha} = \ln 3 \approx 1.099.
\label{eq:lyap}
\end{align}
The Kolmogorov-Sinai (KS) entropy is
\begin{equation}
h_{\rm KS} = \alpha\ln\tfrac{1}{\alpha} + (1-\alpha)\ln\tfrac{1}{1-\alpha} \approx 0.6365.
\label{eq:hks}
\end{equation}
This value is smaller than $h_{\rm KS}(1/2)=\ln 2\approx 0.693$, because the
fraction $\alpha=2/3$ of initial conditions in the weaker-expansion branch
outweighs the stronger one at $1-\alpha=1/3$ in the entropy average.

The fact that the two Lyapunov exponents differ is precisely the reason we choose
$\alpha=2/3$ instead of $\alpha=1/2$. In the symmetric case there is an additional
reflection symmetry $(q,p)\to(1-q,1-p)$ that introduces degeneracies in the quantum
spectrum, complicating statistical studies. The asymmetric map has no such symmetry
and is therefore a cleaner model for eigenvalue statistics.

\subsubsection{Classical openings and survival statistics}

We introduce openings in position space by selecting $M$ of the $N$ channels of
equal width $1/N$ along the $q$-axis. A trajectory is removed from the ensemble
as soon as its $q$-coordinate falls inside an open channel. 
\begin{figure}[ht]
\includegraphics[width=\columnwidth]{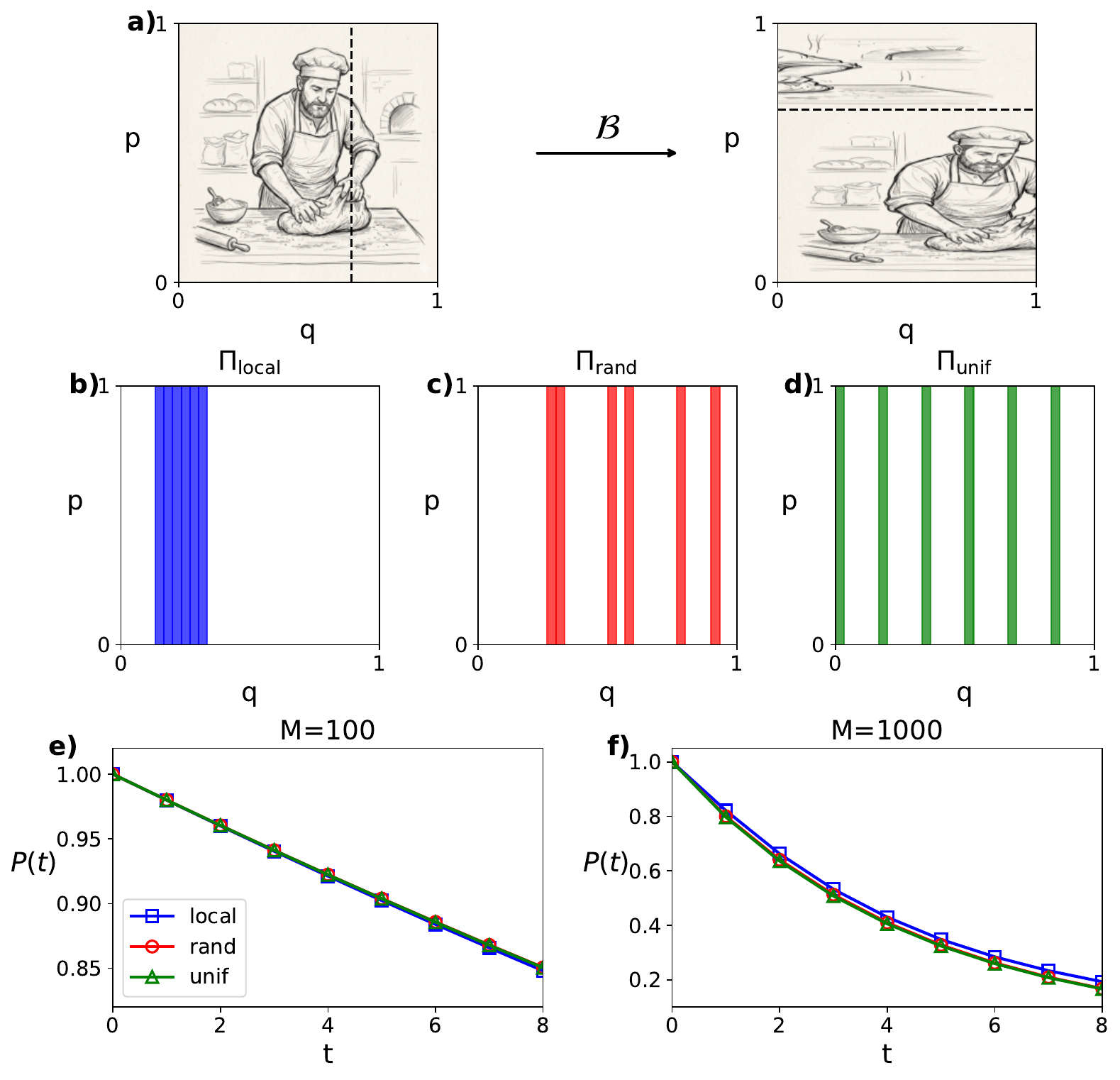}
\caption{%
Representation of the classical transformation, quantum opening configurations,
and classical survival statistics.
(a)~Illustration of the classical asymmetric baker map $\mathcal{B}$. The original
domain (left) is divided by a vertical line at $q=2/3$ (the map's discontinuity).
The transformation stretches and compresses these regions to form the transformed
domain (right), where the boundary is at $p=2/3$.
(b-d)~Schematic of the quantum openings in a phase space of $N=30$ states,
where colored regions indicate the open channels:
(b)~localized ($\Pi_{\rm local}$), (c)~random ($\Pi_{\rm rand}$), (d)~uniform ($\Pi_{\rm unif}$).
(e-f)~Classical survival probability $P(t)$ as a function of time $t$ for $M=100$
(e) and $M=1000$ (f) open channels, with $N=5001$.
Red: random, green: uniform, blue: localized.
Fractal dimensions $d_I$ for $M=100$: $1.968$ (rand, unif, local);
for $M=1000$: $1.649$ (rand), $1.661$ (unif), $1.758$ (local), estimated
via Eq.~\eqref{eq:kantz} with $h_{\rm KS}=0.6365$.%
}
\label{fig:1}
\end{figure}

We study three types
of openings (illustrated in Fig.~\ref{fig:1} for a small phase space):
\begin{itemize}
\item \textit{Localized}: $M$ contiguous channels, with random starting position.
\item \textit{Random}: $M$ channels chosen uniformly at random.
\item \textit{Uniform}: $M$ equispaced channels with step $\lfloor N/M\rfloor$
and a random offset.
\end{itemize}
All results involve averages over many independent realizations of the channel
positions to suppress fluctuations specific to a given configuration.

The survival probability $P(t)$ decays asymptotically as $P(t)\sim C\,e^{-\gamma t}$
where $\gamma$ is the escape rate. The fractal dimension of the repeller
(the set of points that never escape in either time direction) is estimated via
the Kantz-Grassberger relation \cite{altmann13}:
\begin{equation}
d_I = 2 - \frac{\gamma}{h_{\rm KS}}.
\label{eq:kantz}
\end{equation}
In Eq.~\eqref{eq:kantz}, $h_{\rm KS}$ is used as a proxy for the Lyapunov exponent
on the repeller; this is a standard approximation for small to moderate openings
\cite{altmann13}.

Figure~\ref{fig:1}(e,f) shows $P(t)$ for $M=100$ and $M=1000$ open channels out
of $N=5001$ total. For $M=100$ ($M/N\approx 0.02$) all three models give essentially
the same escape rate $\gamma\approx 0.020$ and fractal dimension $d_I\approx 1.97$,
close to the naive estimate $\gamma\approx -\ln(1-M/N)\approx M/N$. In this regime
the opening is small enough that all trajectories mix through it at similar rates
regardless of geometry.

For $M=1000$ ($M/N\approx 0.2$) the three models differ appreciably. The localized
model has a significantly lower escape rate ($\gamma\approx 0.154$, $d_I\approx 1.758$)
compared to random ($\gamma\approx 0.224$, $d_I\approx 1.649$) or uniform
($\gamma\approx 0.216$, $d_I\approx 1.661$). A compact opening is less efficient at
draining probability because it ``blocks'' a contiguous region of $q$-space that
the map can partially avoid through its itinerary structure. These classical
differences will be reflected in the quantum spectral properties at large $M$.

\subsection{Quantum baker map with partial opening}
\label{sec:quantum}

The quantum asymmetric baker map is built following the Saraceno-Voros quantization
procedure \cite{saraceno90,balazvoros89,ermann06baker} with antiperiodic boundary
conditions in both position and momentum (Floquet angles $\theta_q=\theta_p=1/2$).
For a Hilbert space of dimension $N$, with $N=5001=3\times 1667$ ensuring that the
two sub-blocks have integer dimensions $N_1=\lfloor 2N/3\rfloor=3334$ and
$N_2=N-N_1=1667$, the closed propagator is
\begin{equation}
U = F_N \cdot \bigl(F_{N_1}^{-1} \oplus F_{N_2}^{-1}\bigr),
\label{eq:baker_quantum}
\end{equation}
where the discrete Fourier transform matrix with antiperiodic phases has elements
\begin{equation}
(F_d)_{jk} = \frac{1}{\sqrt{d}}\exp\!\left(\frac{2\pi i(j+\frac{1}{2})(k+\frac{1}{2})}{d}\right),
\label{eq:dft}
\end{equation}
with $j,k=0,\ldots,d-1$.
The full propagator $U$ is unitary and its spectrum lies on the unit circle.
The CUE statistics of $U$ in the closed case follow from the fact that the map is
fully chaotic and the asymmetry at $\alpha=2/3$ breaks any time-reversal symmetry
\cite{ermann06baker,haake19}.

The partial opening is implemented by multiplying $M$ columns of $U$ (those
corresponding to the open channels in position) by the amplitude reflectivity $\rho$:
\begin{equation}
U^{(\rho)} = U \cdot \Pi^{(\rho)},
\label{eq:open_quantum}
\end{equation}
where $\Pi^{(\rho)}$ is a diagonal matrix with
\begin{equation}
(\Pi^{(\rho)})_{jj} = \begin{cases} \rho & \text{if channel $j$ is open,} \\ 1 & \text{otherwise.}\end{cases}
\label{eq:projector}
\end{equation}
For $\rho=0$, the opened columns are set to zero (standard projective opening,
equivalent to TCUE). For $\rho=1$, the map is closed. The power reflectivity is
$R=\rho^2$; in the notation of Refs.~\cite{carlo16pre,prado18} our $\rho$ corresponds
to $\rho=\sqrt{R}$.

The eigenvalues $\lambda_i$ of $U^{(\rho)}$ satisfy $|\lambda_i|\le 1$ since
$\Pi^{(\rho)}$ is a contraction for $\rho<1$. The decay rate of the $i$-th state is
$\Gamma_i=-\ln|\lambda_i|$. 
Crucially, for all statistical analysis of spectral fluctuations we retain only
eigenvalues with $|\lambda_i| \ge 0.1$, effectively filtering out
short-lived resonances (completely decayed).
This threshold ensures that the complex spacing ratios are computed over the
relevant part of the spectrum that populates the complex plane. 

An important limiting case is $M=N$ (all channels open): in this case
$\Pi^{(\rho)}=\rho\,\mathbf{1}$ and $U^{(\rho)}=\rho\,U$, so the eigenvalues form a perfect
ring of radius $\rho$ with the phase statistics of the closed CUE. This shows that
the $M=N$ case returns to a quasi-1D spectrum even though $M$ is maximal.

All results presented in this work use $N=5001$, although we have verified that the
statistical behavior (crossover from 1D to 2D spacing statistics) remains
robust for other system sizes, such as $N=10002$. We use averages over 20 realizations 
of the opening geometry for the baker map models, and 50 realizations for PTCUE.

%--------------------------------------------------------------------------
% III. SPECTRAL STATISTICS AND PTCUE
%--------------------------------------------------------------------------
\section{Spectral statistics and the PTCUE}
\label{sec:rmt}

\subsection{The partially truncated CUE}
\label{sec:ptcue}

The partially truncated CUE (PTCUE) is the natural random matrix model for our
system. A PTCUE matrix is obtained from a CUE random matrix $U$ (drawn from the
Haar measure on $\mathrm{U}(N)$) by applying the same partial projector as in
Eq.~\eqref{eq:projector}:
\begin{equation}
U_{\rm PTCUE}^{(\rho)} = U \cdot \Pi^{(\rho)}.
\label{eq:ptcue}
\end{equation}
For $\rho=0$ this is the TCUE \cite{novaes13}, while $\rho=1$ recovers the CUE.
The motivation for this choice is twofold: first, the closed quantum baker map
has CUE statistics \cite{ermann06baker,haake19}; second, the opening procedure
in Eq.~\eqref{eq:open_quantum} is structurally identical to the PTCUE construction.
For a fully chaotic system, universality arguments suggest that all specific features
of the dynamics should be washed out in the spectral statistics for sufficiently
large $M$ and $N$ \cite{haake19}, and the PTCUE should provide the correct prediction.

At $\rho=0$ and $M/N=\mu$ (fixed as $N\to\infty$), the density of eigenvalues of the
TCUE is non-uniform in the complex plane: it peaks near the unit circle
(long-lived states) and decreases toward the center (short-lived states), with
a ``spectral gap'' at $|\lambda|<e^{-\gamma}$ where $\gamma\approx -\ln(1-\mu)$
is the classical escape rate \cite{novaes13}. This is quite different from the
uniform GinUE disk distribution. As $\mu\to 1$ (large opening) the density
approaches the GinUE circular law \cite{santos26}. Increasing $\rho$ above zero shifts
all eigenvalues toward the unit circle: the distribution becomes more concentrated
near $|\lambda|=1$, and in the limit $\rho\to 1$ (closed) all eigenvalues are on the
unit circle.

\subsection{The complex spacing ratio}
\label{sec:z}

A key challenge in studying spectral correlations for non-Unitary operators is
that the eigenvalues are scattered in the complex plane and the usual unfolding
procedure (used for real spectra) does not generalize straightforwardly. 
The complex spacing ratio~\cite{sa20}
\begin{equation}
z_i = \frac{\lambda_{\mathrm{NN},i} - \lambda_i}{\lambda_{\mathrm{NNN},i} - \lambda_i},
\label{eq:z_def}
\end{equation}
where $\lambda_{\mathrm{NN},i}$ and $\lambda_{\mathrm{NNN},i}$ are the nearest
and next-nearest neighbors of $\lambda_i$ in the complex plane, solves this problem
by construction: since the nearest neighbor is by definition closer than the
next-nearest, we have $|z_i|<1$, i.e., $z_i$ lies inside the unit disk.
The ratio $z_i$ effectively performs a local unfolding by using the local level density
as a length scale. Taking the ratio also suppresses the dependence on the
(non-uniform) density of eigenvalues across the disk, making $P(z)$ a genuine
measure of spectral correlations.

The full information about the 2D spectral fluctuations is contained in the joint
distribution $P(|z|,\theta_z)$ where $\theta_z=\arg(z)\in(-\pi,\pi]$. We compute
the 2D histogram in $(|z|,\theta_z)$ space with periodic boundary conditions in
$\theta_z$ and a mild Gaussian smoothing to reduce finite-sampling artifacts.

For two-dimensional Ginibre statistics (GinUE, $\beta_D=2$), the distribution of
$z$ is known analytically \cite{sa20}:
\begin{equation}
p_{\rm GinUE}(z) \propto \frac{|z|^2}{(1+|z|^2)^4},
\label{eq:pginibre}
\end{equation}
where $p$ is the density per unit area in the $z$-plane. The marginal in $|z|$
is $P_{\rm GinUE}(|z|)\propto |z|^3/(1+|z|^2)^4$, which for small $|z|$ behaves
as $|z|^3$ reflecting the level repulsion between nearby eigenvalues.
A numerical fit to $P(|z|)\sim|z|^\beta$ in the range $|z|\in[0,0.75]$ gives an
effective exponent $\beta\approx 2.5$ for the GinUE distribution (slightly less
than 3 because the $(1+|z|^2)^4$ denominator reduces the apparent slope in this
range). For the GinUE the phase distribution $P(\theta_z)$ is nearly uniform with
a small dip near $\theta_z=0$, arising because the nearest and next-nearest
neighbors are rarely in the same direction.

For a purely 1D spectrum (eigenvalues on a curve or ring), the ratio $z$ is real
and positive (or negative), so $\theta_z$ is concentrated at $0$ or $\pm\pi$
depending on whether the NN and NNN are on the same or opposite sides of $\lambda_i$.
Correspondingly, $P(|z|)$ is approximately flat for small $|z|$ (no level repulsion
in 1D in the $z$ variable), i.e., $\beta\approx 0$.

Two geometrically natural regions have zero probability density regardless of the
statistics. First, $|z|\to 0$ corresponds to the nearest neighbor being essentially
on top of $\lambda_i$ (``near-degeneracy''), which is suppressed by level repulsion
in any repelling ensemble. Second, $|z|\to 1$ combined with $\theta_z\to 0$
corresponds to nearest and next-nearest neighbors being at the same distance from
$\lambda_i$ and in the same direction: this would mean the three points are
collinear with equal spacing, which is also improbable and excluded by definition
for the nearest neighbor (the NNN is always farther than the NN).

\subsection{Eigenvalue and $z$ distributions: representative cases}
\label{sec:rep_cases}

\begin{figure*}[ht]
\includegraphics[width=0.82\textwidth]{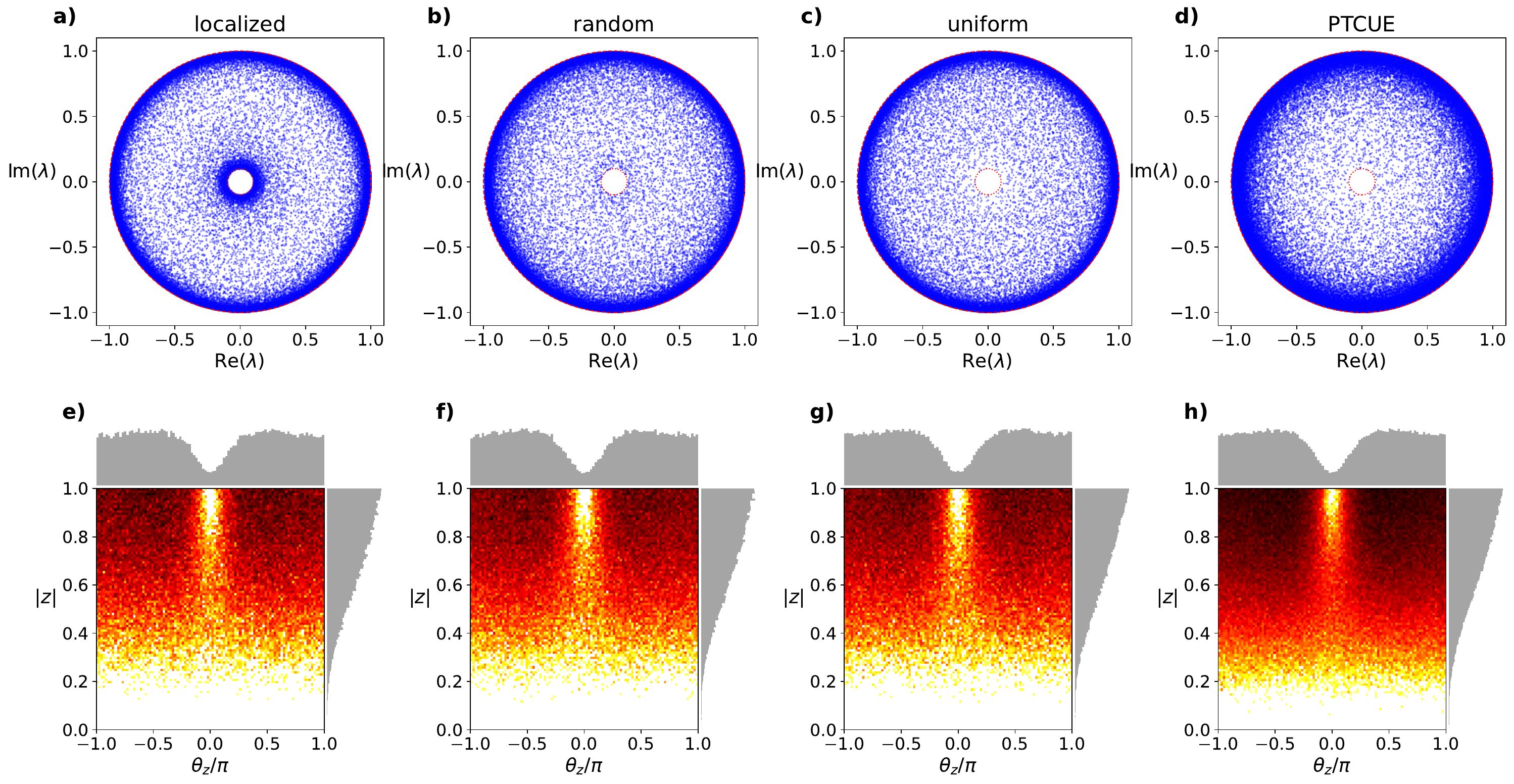}\\[2pt]
\includegraphics[width=0.82\textwidth]{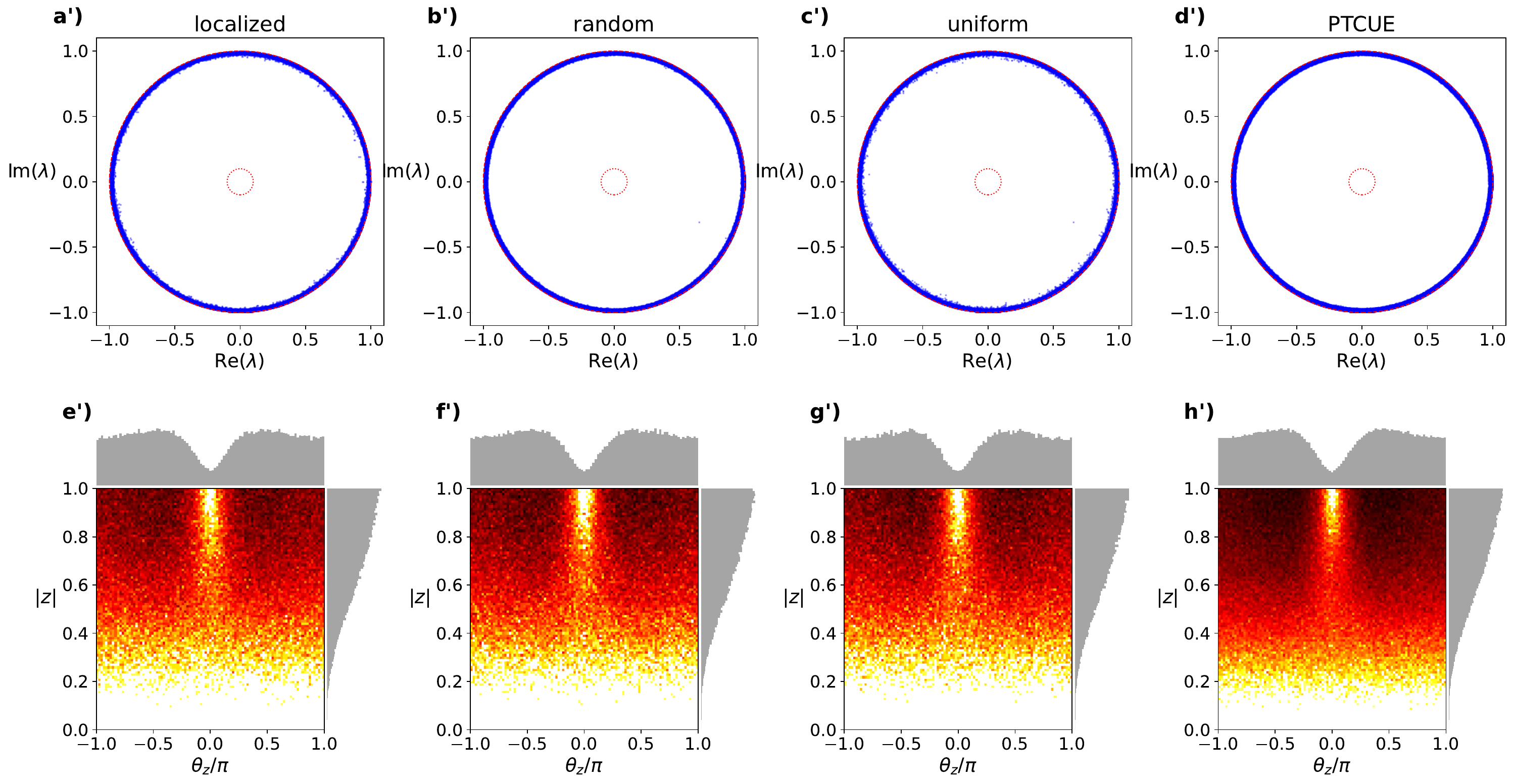}
\caption{%
Spectral properties and complex spacing ratio distributions for $M=100$.
Top block ($\rho=0$, fully open): rows (a-d) show the eigenvalue distribution
in the complex plane for localized~(a), random~(b), uniform~(c), and PTCUE~(d);
rows~(e-h) show the joint density of $z$ in polar coordinates
$(\theta_z/\pi,|z|)$ on a log color scale, with marginal distributions.
Bottom block ($\rho=0.5$, partial opening): same layout~(a$'$-d$'$) and (e$'$-h$'$).
In all panels the red dashed circle marks the unit circle and the dotted circle
marks the filter threshold $|\lambda|=0.1$.
Data from 20 realizations (baker models) and 50 (PTCUE).%
}
\label{fig:2}
\end{figure*}

Figure~\ref{fig:2} displays the eigenvalue spectra for the three asymmetric 
baker map opening models ("localized", "random", and "uniform") alongside 
the PTCUE. The top row (panels a--d) shows the results for $\rho=0$ 
(fully open), while the third row (panels a'--d') corresponds to
$\rho=0.5$, both for $M=100$ open channels. As expected from their 
different decay times shown in Fig.~\ref{fig:1}, the three models exhibit 
distinct eigenvalue distributions in the radial coordinate $|\lambda|$. 
For the partially open case ($\rho=0.5$), the eigenvalues cluster 
closer to the unit circle for $M=100$, a feature that was
previously observed only for a very small number of open channels.
Despite these macroscopic differences in the spectra, the 
corresponding joint distributions of $(\theta_z/\pi, |z|)$ 
for the three models, at both $\rho=0$ and $\rho=0.5$, 
appear to have converged to a typical shape that we will 
refer to as the \textit{baseline} distribution. 
This two-dimensional representation is highly valuable for 
understanding the unfolded level repulsion. Specifically, the 
baseline distribution is characterized by a power-law growth 
in $|z|$ with an exponent $\beta \approx 2.5$ (arising from 
level repulsion), a predominantly uniform angular distribution, 
and a pronounced dip near $\theta_z \approx 0$ at large moduli, 
which is another consequence of level repulsion. The PTCUE 
results display a similar baseline behavior, albeit with 
slight differences and better statistics. Crucially, 
this shows that the usual or \textit{baseline} $z$ distribution 
can be robustly observed even when the spectrum exhibits a narrow 
radial width.

%--------------------------------------------------------------------------
% IV. EVOLUTION OF SPECTRAL STATISTICS WITH M AND rho
%--------------------------------------------------------------------------
\section{From quasi-1D to 2D: evolution with $M$ and $\rho$}
\label{sec:evol}

\subsection{Mean and standard deviation of $|\lambda|$}
\label{sec:mean_std}

\begin{figure}[ht]
\includegraphics[width=\columnwidth]{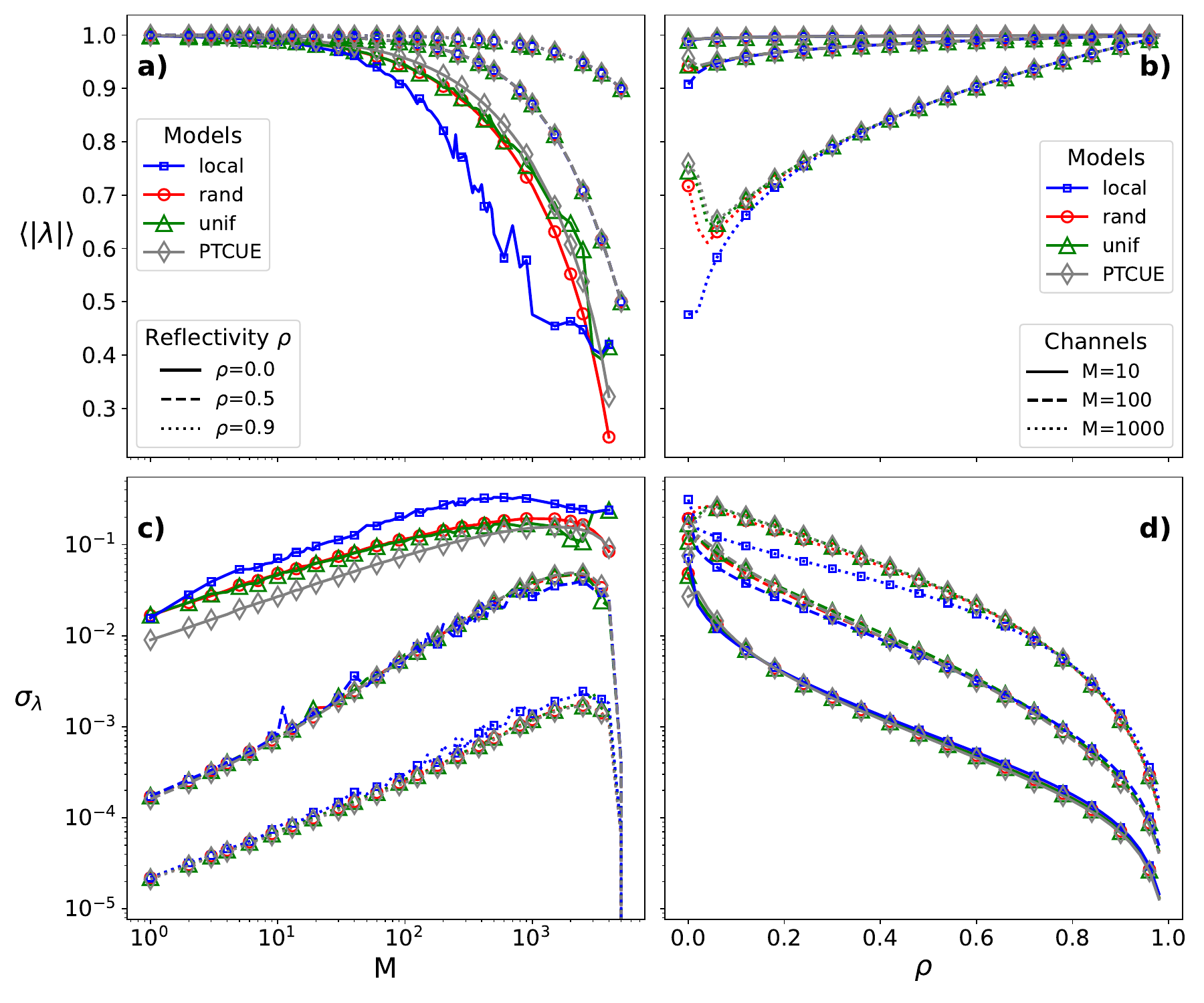}
\caption{%
Statistical properties of eigenvalues with $|\lambda|\ge 0.1$ for the asymmetric
open baker map with $N=5001$.
(a)~Mean $\langle|\lambda|\rangle$ vs.\ $M$ (log scale).
(b)~Mean $\langle|\lambda|\rangle$ vs.\ $\rho$ (linear scale).
(c)~Standard deviation $\sigma(|\lambda|)$ vs.\ $M$ (log-log scale).
(d)~Standard deviation $\sigma(|\lambda|)$ vs.\ $\rho$ (log-linear).
Models: local (blue squares), rand (red circles), unif (green
triangles), PTCUE (gray diamonds).%
}
\label{fig:3}
\end{figure}

Figure~\ref{fig:3} shows the mean $\langle|\lambda|\rangle$ 
(top row, panels a and b) and the standard deviation $\sigma(|\lambda|)$ 
of the eigenvalue moduli as functions of the number of open channels $M$ 
and the reflectivity parameter $\rho$. The results are presented for the 
three baker map models and the PTCUE. The top row illustrates where the bulk
of the spectrum is radially centered. For small openings (i.e., a small 
number of channels, $M \to 1$), the spectrum behaves almost as a unitary 
ring, with $\langle|\lambda|\rangle$ approaching 1. This clustering is 
also reflected in the standard deviation, which becomes very small for 
$M$ values close to 1. Similarly, it is clear that as the reflectivity 
$\rho$ increases, the mean eigenvalue moduli consistently converge toward 
the unit circle. While there are subtle differences between the three baker
map models and the PTCUE, their overall trends are largely similar. 
One of the most prominent deviations is observed in the mean value 
of $|\lambda|$ for the fully open case ($\rho=0$). Under these conditions, 
the eigenvalue moduli are systematically lower. Consequently, 
the applied numerical filter---which discards eigenvalues with 
$|\lambda| < 0.1$---significantly reduces the available statistics. 
It is important to note that for each value of $M$ and $\rho$, the data 
is averaged over 20 random opening realizations for each baker map model, 
and 50 realizations for the PTCUE. Because of the aforementioned filtering 
and the overall low moduli at $\rho=0$, the data for the extreme case 
of $\rho=0$ and large channel numbers (e.g., $M=1000$) exhibits 
considerable statistical noise, as very few eigenvalues survive 
the $|\lambda| > 0.1$ cutoff.

\subsection{Power-law exponent $\beta$ of $P(|z|)$}
\label{sec:beta}

\begin{figure}[ht]
\includegraphics[width=\columnwidth]{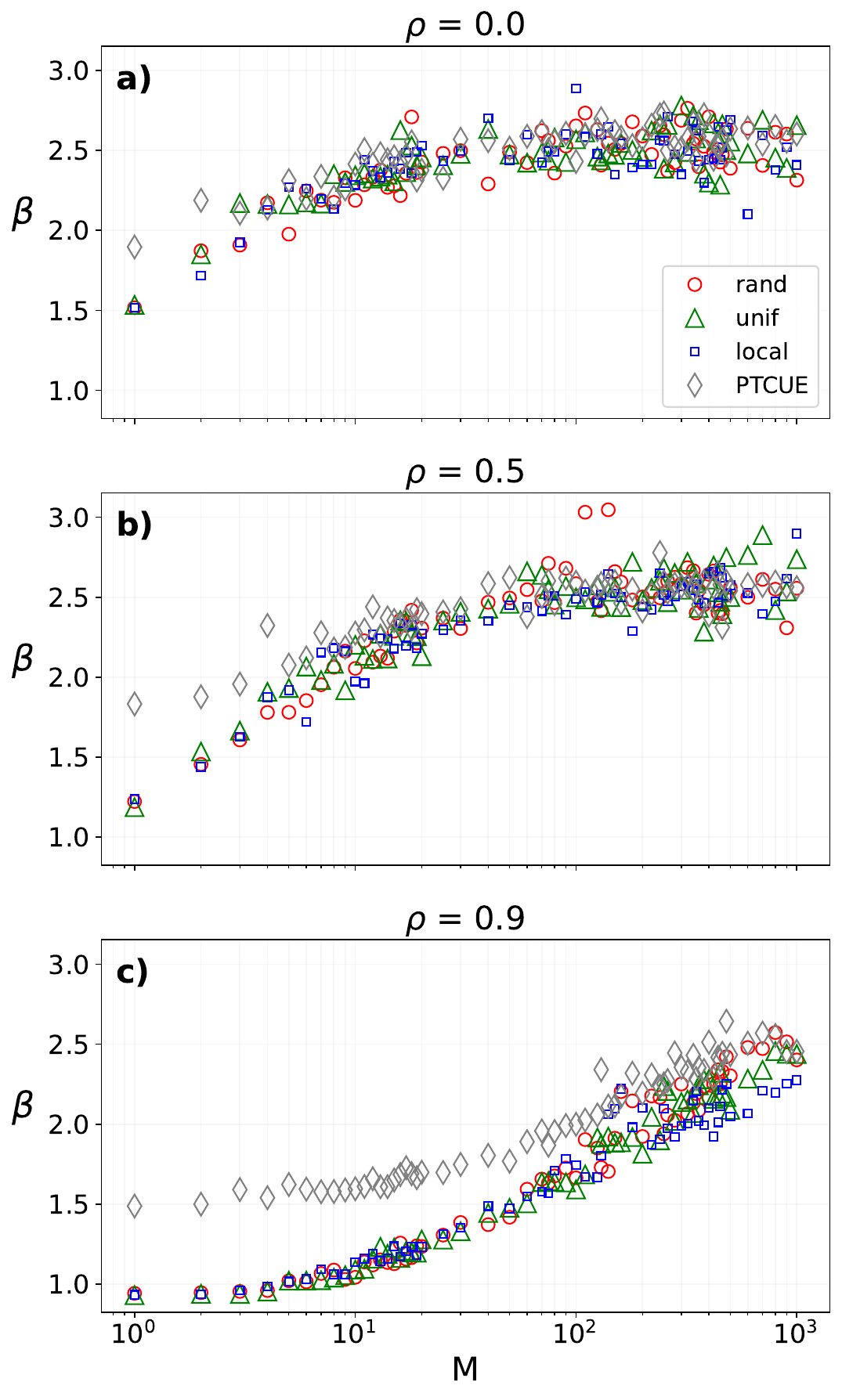}
\caption{%
Power-law exponent $\beta$ of $P(|z|)\sim|z|^\beta$ as a function of $M$ (log scale)
for opening strengths (a)~$\rho=0$, (b)~$\rho=0.5$, (c)~$\rho=0.9$.
Four models are compared in each panel: rand (red circles), unif (green
triangles), local (blue squares), PTCUE (gray diamonds). The exponent is obtained
by a linear regression in log-log scale for $|z|\in[0,0.75]$.%
}
\label{fig:4}
\end{figure}

\begin{figure*}[tbp]
\includegraphics[width=\textwidth]{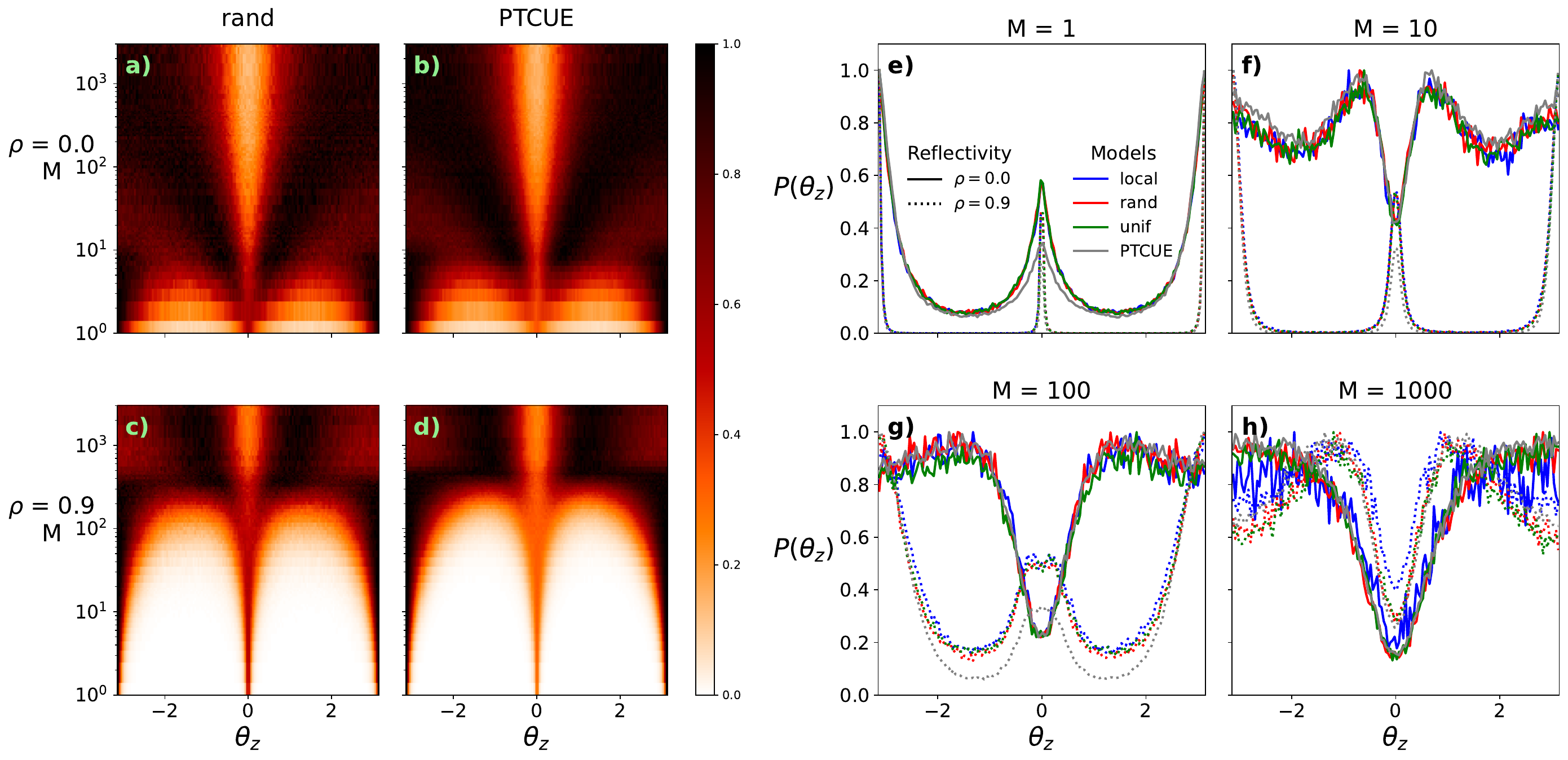}
\caption{%
Phase distribution $P(\theta_z)$ of the complex spacing ratio $z$ for the open
baker map with $N=5001$.
Panels (a-d): heatmaps of the density of $\theta_z$ as a function of $M$ (log
scale) for the random model and PTCUE at $\rho=0$ (top) and $\rho=0.9$ (bottom).
To facilitate visualization across different regimes and avoid saturation in
localized regions, the relative density for each $M$ is normalized to its maximum
value rather than setting the total integral to unity.
Panels (e-h): direct comparison of $P(\theta_z)$ for specific $M$ values
($M=1$, $10$, $100$, $1000$). Four models are shown: local (blue), rand (red),
unif (green), PTCUE (gray). Solid lines: $\rho=0$; dashed lines: $\rho=0.9$.%
}
\label{fig:5}
\end{figure*}

Figure~\ref{fig:4} shows the power-law exponent $\beta$, extracted 
from the relation $P(|z|) \sim |z|^\beta$, as a function of the 
number of open channels $M$ on a logarithmic scale. 
The results are presented for reflectivities $\rho=0$, $0.5$, 
and $0.9$ (panels a, b, and c, respectively), comparing the three
baker map models with the PTCUE. For the fully open case 
($\rho=0$, panel a), the exponent $\beta$ at the smallest 
opening size ($M=1$) takes values between 1.5 and 2. As $M$ 
increases, $\beta$ grows monotonically, converging toward 
the expected value of $\beta \approx 2.5$, which is consistent 
with the 2D Ginibre ensemble (GinUE) prediction. Notably, for 
small values of $M$, the PTCUE yields systematically higher 
$\beta$ values than the three deterministic models, indicating 
that the random matrix model converges more rapidly to the 
Ginibre limit.In panel b, which corresponds to a partial 
reflectivity of $\rho=0.5$, the system exhibits a trend 
similar to the fully open case, albeit with slightly lower 
$\beta$ values across the board. The exponent still converges 
to the expected 2.5 limit, but this saturation is delayed, 
requiring a larger number of open channels $M$ to be reached.
The strong partial reflectivity case ($\rho=0.9$, panel c) 
displays a distinctly different behavior. Here, the values 
of $\beta$ remain below 2 for a wide range of channel numbers 
up to $M \sim 100$. Convergence to the 2.5 limit is only achieved 
at very large openings, on the order of $M \sim 1000$. 
Furthermore, it is particularly notable in this regime 
that the $\beta$ values for the PTCUE are significantly 
higher than those of the three baker map models, 
highlighting a pronounced discrepancy between the 
deterministic maps and the random matrix model under 
near-closed conditions. 
Ultimately, these panels illustrate how both the 
opening size $M$ and the partial reflectivity $\rho$ 
jointly control the spectral shape and the convergence to 
2D Ginibre statistics. 
Increasing $\rho$ effectively slows down this crossover, shifting 
the saturation of $\beta$ to much higher channel numbers.

\subsection{Phase distribution of $z$ and the 2D crossover}
\label{sec:theta}

Figure~\ref{fig:5} illustrates the evolution of the phase distribution 
$P(\theta_z)$ as a function of the number of open channels $M$.
Panels (a) through (d) display $P(\theta_z)$ (horizontal axis) 
versus $M$ (vertical axis, logarithmic scale) as heatmaps. 
Panels (a) and (b) show the distributions for the random baker 
map model and the PTCUE, respectively, at $\rho=0$. 
The other two deterministic models exhibit behavior very 
similar to the random case. Panels (c) and (d) show the 
corresponding results for $\rho=0.9$. At $M=1$, the spectrum 
is in a quasi-unitary (quasi-1D) regime, resulting in three distinct 
peaks at $\theta_z = -\pi, 0,$ and $\pi$. As $M$ grows, 
the distribution undergoes a profound transition from quasi-1D to 2D, 
converging toward a roughly uniform distribution with a pronounced
dip near $\theta_z \approx 0$ due to level repulsion.
The stark differences that characterize this transition are most evident in the evolution 
of the central feature at $\theta_z=0$, which transforms from a global
maximum (at small $M$ and large $\rho$) into a minimum (at large $M$). 
This crossover unfolds smoothly: the central maximum at $\theta_z=0$ 
splits into two symmetric peaks, a consequence of the parity symmetry
of $\theta_z$. As $M$ increases, these two maxima drift outward toward 
larger $|\theta_z|$ values while simultaneously broadening and losing amplitude. 
Eventually, they smear out into a uniform distribution for
$|\theta_z| > \pi/2$, leaving behind the characteristic 
level-repulsion minimum at $\theta_z=0$. This dynamic is 
further detailed in panels (e--h), which show cross-sections of $P(\theta_z)$ 
for all three baker map models and the PTCUE at specific channel numbers
$M=1, 10, 100,$ and $1000$. Comparing the fully open ($\rho=0$, solid lines) 
and highly reflective ($\rho=0.9$, dashed lines) cases reveals 
significant structural differences. For $\rho=0.9$, the 
peaks at $-\pi, 0,$ and $\pi$ are considerably narrower, creating 
intervening regions of strictly zero probability.
While this strictly zero probability is a direct 
signature of a strongly quasi-1D spectrum, it is not observed in the 
$\rho=0$ case. Finally, the delayed transition for strong reflectivities 
is clearly seen by examining panel (g) for $M=100$: at $\rho=0.9$, 
the maximum at $\theta_z=0$ is still present, yet its shape and width 
are distinctly different from the corresponding maximum seen in panel 
(e) for $M=1$ and $\rho=0$.

%--------------------------------------------------------------------------
% V. COMPARING THE THREE MODELS
%--------------------------------------------------------------------------
\section{Model comparison via Bhattacharyya overlap}
\label{sec:compare}

To quantify the similarity between the 2D $z$ distributions of different models,
we use the Bhattacharyya coefficient \cite{bhattacharyya43}, which for two
probability distributions $P_1$ and $P_2$ on a common domain is defined as:
\begin{equation}
\mathcal{S}(P_1,P_2) = \int \sqrt{P_1(\mathbf{x})\, P_2(\mathbf{x})}\, d^2x,
\label{eq:bhatta}
\end{equation}
where $\mathbf{x}=(\abs{z},\theta_z)$ and the integral is over the domain
$\abs{z}\in[0,1]$, $\theta_z\in(-\pi,\pi]$. Both distributions are normalized
to unit integral before computing $\mathcal{S}$, and discretized on an
$80\times 80$ grid. By the Cauchy-Schwarz inequality $0\le\mathcal{S}\le 1$,
with $\mathcal{S}=1$ if and only if the distributions are identical. This metric is well
suited for comparing 2D distributions because it is sensitive to both the
shape and the tails of the distributions.

\begin{figure}[tbp]
\includegraphics[width=\columnwidth]{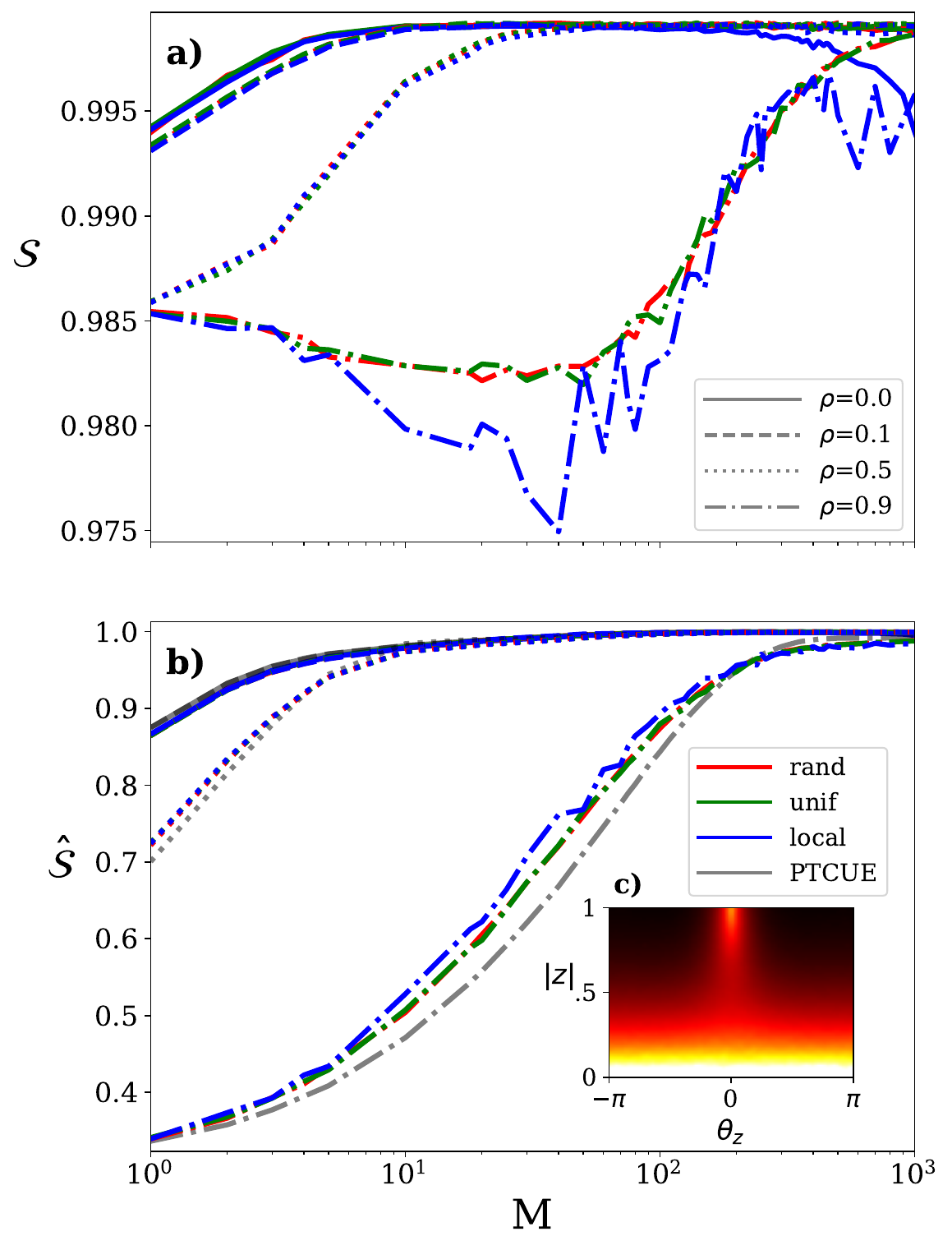}
\caption{%
Statistical similarity and convergence of complex spacing distributions, measured
by the Bhattacharyya coefficient $\mathcal{S}$ [Eq.~\eqref{eq:bhatta}] computed
over the 2D $z$ distribution in $(\abs{z},\theta_z)$ space.
(a)~Similarity $\mathcal{S}$ between each baker map model and PTCUE for the same
$\rho$. Three values of $\rho$ are compared (see legend). At large $M$ all models
converge to PTCUE.
(b)~Convergence $\hat{\mathcal{S}}$ of all models (at $\rho=0$) toward the base
distribution $P_{\rm base}$, constructed from PTCUE with $\rho=0$ averaged over
$M\in[100,2000]$.
(c)~Inset: the base distribution $P_{\rm base}$ in the $(\abs{z},\theta_z)$ plane,
representing the 2D large-opening limit.%
}
\label{fig:6}
\end{figure}

Figure~\ref{fig:6} analyzes the quantitative agreement between 
the distributions using a similarity metric.
Figure~\ref{fig:6}(a) shows the similarity $\mathcal{S}$ between each of the three baker
map models and the corresponding PTCUE (evaluated at the same $M$ and $\rho$) 
as a function of the number of open channels $M$ on a logarithmic scale. 
The curves are shown for four representative reflectivities:
$r=0.0, 0.1, 0.5,$ and $0.9$. Overall, the deterministic models 
and the random matrix predictions are highly similar, yielding metric 
values above $\mathcal{S} > 0.975$. However, for small openings 
($M \sim 1$), the convergence of the models toward the PTCUE is not 
perfectly exact, which echoes the slight discrepancies previously 
observed in the power-law exponents in Fig.~\ref{fig:4}. 
As $M$ increases, this similarity steadily improves. Crucially, 
the larger the reflectivity $\rho$, the larger the initial discrepancy, 
meaning the system requires a significantly higher number of open channels 
$M$ to fully converge to $\mathcal{S} \approx 1$.

Figure~\ref{fig:6}(b) 
employs a related metric, $\hat{\mathcal{S}}$, to evaluate the convergence 
of all models (as well as the PTCUE itself) toward a universal ``base'' 
distribution, $P_{\rm base}$. This reference base distribution, 
visualized in the inset (panel c), is defined by averaging the PTCUE at 
$r=0.0$ over large values of $M$. In this asymptotic 2D limit, 
the level repulsion is completely clear, and the characteristic minimum 
at $\theta_z=0$ is fully developed. When comparing the models to this 
$P_{\rm base}$, there is a substantial difference at small values of $M$, 
reflecting the quasi-1D nature of the spectrum in that regime. 
As $M$ grows, $\hat{\mathcal{S}}$ increases monotonically. 
For small values of $\rho$, the transition is relatively fast, 
and convergence to the base distribution is achieved at smaller 
values of $M$. Conversely, for strong partial reflectivities such as 
$\rho=0.9$, the crossover is dramatically delayed; the metric only 
approaches $\hat{\mathcal{S}} \approx 1$ at exceptionally large openings, 
on the order of $M \sim 1000$. Together, these panels confirm that the PTCUE 
is an exceptionally robust model for all three opening geometries. 
The transition from a quasi-1D spectrum to the fully developed 2D base 
distribution is a smooth, continuous process controlled jointly by $M$ and $\rho$.

\begin{figure}[tbp]
\includegraphics[width=\columnwidth]{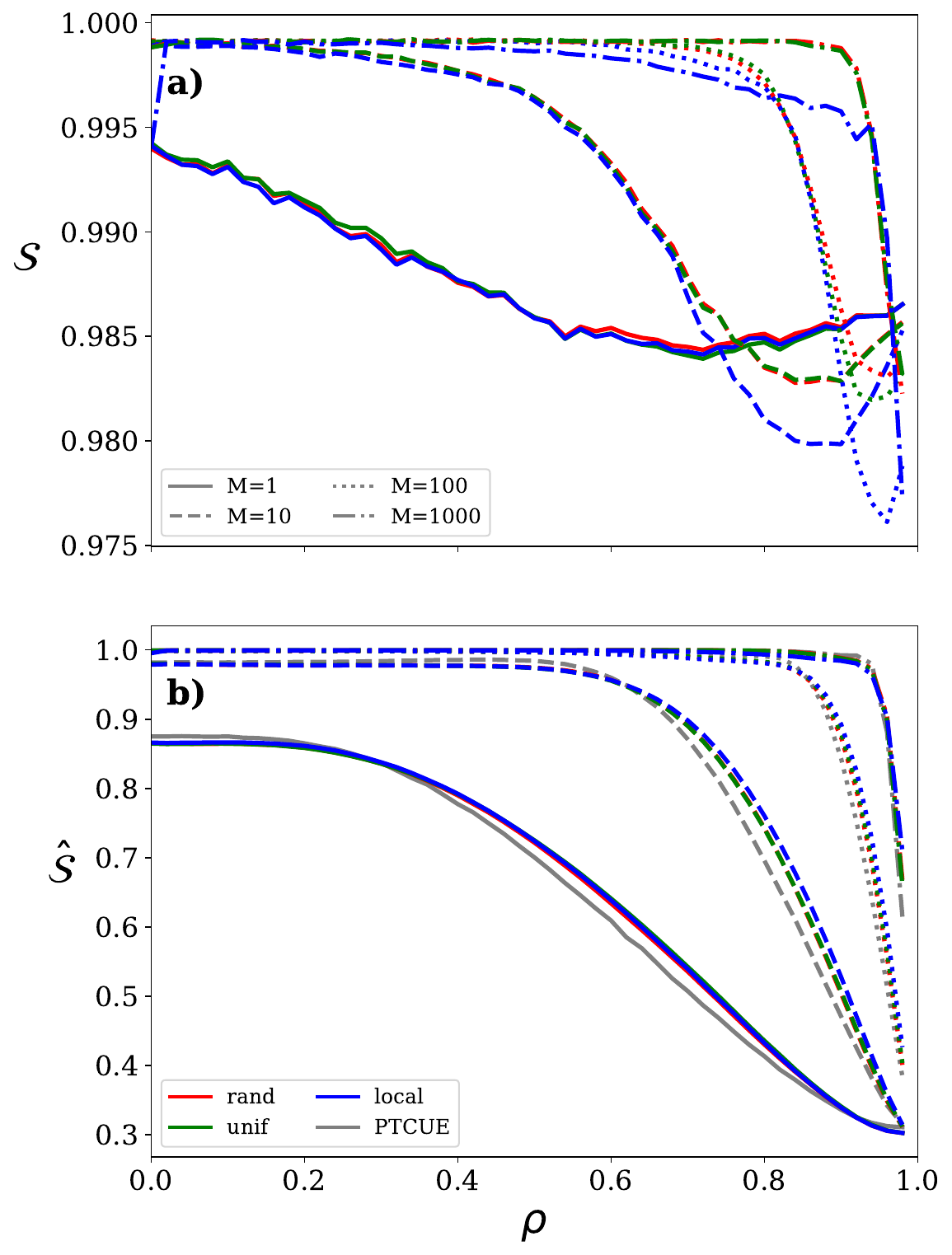}
\caption{%
Statistical similarity and convergence of complex spacing distributions 
as a function of the amplitude reflectivity $\rho$ for representative opening 
sizes $M=1, 10, 100,$ and $1000$.
(a)~Bhattacharyya coefficient $\mathcal{S}$ between each baker map model 
and the PTCUE benchmark evaluated at the same $(M,\rho)$.
(b)~Convergence $\hat{\mathcal{S}}$ of the deterministic models and the 
PTCUE toward the universal base distribution $P_{\rm base}$ [shown in Fig.~\ref{fig:6}(c)].%
}
\label{fig:7}
\end{figure}

To complement this analysis, Fig.~\ref{fig:7} explores the behavior of 
the similarity metrics from a complementary perspective, displaying the 
results explicitly as a function of the amplitude reflectivity $\rho$ for 
four representative opening sizes, $M=1, 10, 100,$ and $1000$, 
capturing different regimes of openness. 

Figure~\ref{fig:7}(a) presents the direct comparison between the
joint $(\abs{z},\theta_z)$ distributions of the three baker map 
models and the PTCUE benchmark. In general, the deviations from 
the random matrix predictions remain remarkably small across 
almost the entire range of parameters. However, the discrepancies 
become more visible for small opening sizes (represented by solid 
lines) and in the limit of strong partial reflectivity ($\rho \to 1$).
 Notably, as the number of open channels $M$ increases, these 
 localized discrepancies are progressively suppressed, and one
  must approach closer to the closed limit ($\rho \approx 1$) to 
  detect any meaningful breakdown of the universal PTCUE behavior.

Figure~\ref{fig:7}(b) shows the convergence metric $\hat{\mathcal{S}}$ 
when evaluating the models against the universal base distribution 
$P_{\rm base}$ [previously introduced in Fig.~\ref{fig:6}(c)]. For 
all cases, $\hat{\mathcal{S}}$ decreases monotonically as the 
reflectivity $\rho$ increases. For the single-channel case ($M=1$), 
the spectrum remains structurally constrained, and $\hat{\mathcal{S}}$ 
never exceeds a value of $0.9$, emphasizing that a minimal opening 
cannot drive the system fully into the asymptotic 2D regime. As $M$ 
grows, the curves shift upward, showing values of $\hat{\mathcal{S}}$ 
much closer to unity for low reflectivities. In these cases, a 
significantly larger value of $\rho$ is required to trigger the 
drop in similarity, reflecting how a large number of channels 
effectively sustains the 2D character of the spectrum until the 
system is nearly closed.
%--------------------------------------------------------------------------
% VI. CONCLUSIONS
%--------------------------------------------------------------------------
\section{Conclusions}
\label{sec:conclusions}

We have studied the complex eigenvalue statistics of the asymmetric quantum baker
map with partial projective openings, using the complex spacing ratio $z$ as the
main observable. The combination of a fully chaotic classical limit, a clean
quantization, three geometrically distinct opening types, and a continuous partial
reflectivity parameter $\rho$ makes this system an ideal laboratory for studying the
crossover between different spectral regimes.

The main finding is a smooth crossover between a quasi-1D regime and a 2D
Ginibre-like regime, controlled by both $M$ (number of open channels) and $\rho$
(amplitude reflectivity). In the quasi-1D regime ($M$ small or $\rho$ large), the
eigenvalue spectrum is quasi-unitary and the $z$ statistics reflect the nearly
1D arrangement of eigenvalues near the unit circle: $P(\theta_z)$ is peaked at
$\theta_z=0,\pm\pi$ and the exponent $\beta$ in $P(|z|)\sim|z|^\beta$ is near
zero. As $M$ grows (or $\rho$ decreases), the spectrum fills the complex disk and
the statistics evolve toward the 2D GinUE limit with nearly uniform $P(\theta_z)$
and $\beta\approx 2.5$.

The partial reflectivity $\rho$ provides a universal control of this crossover.
What was previously described as a ``quantum opening'' regime (very small
classical opening, quasi-1D statistics for $\rho=0$) is now understood as a
smooth function of both $M$ and $\rho$: for $\rho=0.9$ and $M=100$, the spectrum
is as quasi-1D as for $\rho=0$ and $M=1$. This has an important consequence:
the transition to 2D statistics is not a singularity at small opening size
but rather a continuous crossover that can be accessed by tuning $\rho$ from
any moderate $M$. This is directly relevant for experiments where the partial
reflectivity of a cavity or billiard can be controlled.

The three opening models (localized, random, uniform) converge to the PTCUE
prediction at large $M$ with good quantitative agreement ($\mathcal{S}>0.95$
for $M\gtrsim 100$, $\rho=0$). Differences are largest at small $M$ and large $\rho$,
where the specific geometry of the opening affects how efficiently the classical
trajectories are absorbed. The localized model shows the slowest convergence,
consistent with its lower classical escape rate and larger repeller dimension.
Nevertheless, even for the localized model the crossover is smooth and continuous.

These results also have implications for the study of spectral correlations
in other open maps. While recent work on the leaky standard map \cite{santos26}
highlighted the role of dynamical symmetries (AI$^\dagger$ class) in shaping
short-range correlations for localized leaks, our asymmetric baker map
effectively removes those symmetries by construction. This allows for a clean
demonstration of the universal convergence to the PTCUE circular law, confirming
that the 1D-to-2D crossover is a robust feature of chaotic resonance spectra
regardless of the underlying symmetry constraints.

The crossover we observe shows no evidence of an abrupt transition, either as
a function of $M$ or of $\rho$. This is in contrast to what one might naively
expect from the dramatic difference between quasi-1D and 2D statistics, but is
consistent with the general expectation that smooth changes in the system
parameters should lead to smooth changes in the spectrum. Whether a sharper
transition exists in the $N\to\infty$ limit for fixed $M/N$ and $\rho$ remains
an interesting open question.

Finally, we note that studying the full 2D joint distribution $P(|z|,\theta_z)$
rather than marginals alone captures information that marginals miss. In particular,
the forbidden region near ($|z|\to 1$, $\theta_z\to 0$) in the 2D plot and
the evolution of the maximum of $P(\theta_z)$ with $M$ are features that are
invisible in the 1D marginals but provide a clear picture of the crossover.
We expect these results to motivate further theoretical work on the distribution
of $z$ for the PTCUE and for other non-Unitary random matrix ensembles with
partial truncations. One interesting avenue we are currently pursuing is 
uncovering effective analytical expressions that could describe this crossover, 
which we conjeture to be universal.

\begin{acknowledgments}
This work was supported by CONICET and CNEA (Argentina). 
Specifically, the authors acknowledge financial support 
from CONICET under Grant No.~PIP 2022-2024 GI - 11220210100208CO. 
\end{acknowledgments}

%--------------------------------------------------------------------------
% FIGURES
%--------------------------------------------------------------------------
% BIBLIOGRAPHY
%--------------------------------------------------------------------------
\bibliography{paper}

\end{document}